\documentclass[preprint,12pt,a4paper]{elsarticle}

\usepackage[T1]{fontenc}
\usepackage{amssymb}
\usepackage{amsmath}

\usepackage{float}
\restylefloat{table}

\usepackage[hyphens]{xurl}
\usepackage{hyperref}
\usepackage[hyphenbreaks]{breakurl}

\usepackage{listings}
\usepackage{color}

\definecolor{darkblue}{rgb}{0,0,.6}
\definecolor{darkred}{rgb}{.6,0,0}
\definecolor{darkgreen}{rgb}{0,.6,0}
\definecolor{red}{rgb}{.98,0,0}

\journal{SoftwareX}

\newcommand{\myemail}{igor.s.krivenko@gmail.com}
\newcommand{\libcommute}{{\it libcommute}}
\newcommand{\pycommute}{{\it pycommute}}

\begin{document}

\begin{frontmatter}

\title{libcommute/pycommute: A quantum operator algebra domain-specific language and exact diagonalization toolkit}

\author{I. Krivenko}
\ead{ikrivenk@physnet.uni-hamburg.de}
\address{I. Institut f\"ur Theoretische Physik, Universit\"at Hamburg, Notkestra\ss e 9, 22607 Hamburg, Germany}

\begin{abstract}
I present \libcommute, a C++11/14/17 template library that implements a domain-specific language for easy manipulating of polynomial operators used in the quantum many-body theory, as well as a software development toolkit for exact diagonalization codes. The library is written with expressiveness, extensibility and performance in mind. It features simple syntax for commonly used abstractions and algorithms, is well documented and covered by unit tests.
\libcommute\ is supplemented with Python 3 bindings called \pycommute. They are useful for solving small scale diagonalization problems, rapid prototyping and wrapping of high performance \libcommute-based computational cores in Python.
\end{abstract}

\begin{keyword}
quantum \sep algebra \sep many body physics \sep exact diagonalization
\end{keyword}

\end{frontmatter}

\section*{Required Metadata}
\label{sec:metadata}

\section*{Current code version}
\label{sec:version}

\begin{table}[H]
\begin{tabular}{|l|p{6.5cm}|p{6.5cm}|}
\hline
\textbf{Nr.} & \textbf{Code metadata description} & \textbf{Please fill in this column} \\
\hline
C1 & Current code version & v0.7.0 \\
\hline
C2 & Permanent link to repository used for this code version & 
\begin{itemize}
    \item \libcommute:
        \url{https://github.com/krivenko/libcommute/releases/tag/v0.7.0}
    \item \pycommute:
        \url{https://github.com/krivenko/pycommute/releases/tag/v0.7.0}
\end{itemize}
 \\
\hline
C3 & Code Ocean compute capsule & \url{https://codeocean.com/capsule/9762614/tree} \\
\hline
C4 & Legal Code License & Mozilla Public License 2.0 \\
\hline
C5 & Code versioning system used & git \\
\hline
C6 & Software code languages, tools, and services used & C++11/17, CMake, Python 3.6+ \\
\hline
C7 & Compilation requirements, operating environments \& dependencies & 
A C++11-compatible compiler (C++17 support is required for some features and to build \pycommute), pybind11 2.6.0+, NumPy, Sphinx 2.1.0+ to build documentation.
\\
\hline
C8 & Link to developer documentation & 
\begin{itemize}
    \item \libcommute: \url{https://krivenko.github.io/libcommute/}
    \item \pycommute: \url{https://krivenko.github.io/pycommute/}
\end{itemize} \\
\hline
C9 & Support email for questions & \myemail \\
\hline
\end{tabular}
\caption{Code metadata}
\label{} 
\end{table}


\section{Motivation and significance}
\label{sec:motivation}

The last three decades have seen a boom in the field of computational quantum many-body theory. It has been driven by the limited efficiency of analytical approaches on the one hand, and by the ever growing availability of computer resources on the other.
Exact diagonalization (ED) \cite{EDBook,LanczosChapter} has become one of the most important and commonly used families of numerical methods in the field. In a nutshell, an ED calculation amounts to reducing a quantum system of interest to a finite effective model, which retains some essential features of the original system, and whose Hamiltonian is amenable to numerical diagonalization. Computed energy levels and eigenstates are then combined to form expectation values of physical observables and correlation functions. In addition to being a stand-alone solver, ED is an integral part of more sophisticated methods, for instance the Hybridization Expansion Quantum Monte Carlo \cite{WernerCTHYB} and master equation solvers for non-equilibrium dynamics \cite{LindbladSolution}.
In many cases, ED can provide physical insights without revealing the full eigensystem of the problem. It is often sufficient to perform partial diagonalization within a sector defined by a set of quantum numbers, or to apply an iterative method \cite{PowerIteration,ARPACK,PRIMME} to compute only the ground and a few excited states.

In this paper, I present \libcommute\ \cite{libcommute}, a C++ library that aims at establishing a reliable and highly extensible framework for developing ED codes. Development of \libcommute\ has been pursuing the following goals.
\begin{itemize}
    \item Providing a lightweight, header-only library without external dependencies written in the well-established C++11 dialect (some advanced features require the more recent C++17).
    
    \item Employing C++11 meta-programming techniques to ensure expressiveness, ease of use, flexibility and a high degree of extensibility.
    
    \item Orientation towards high performance computing and solution of very large scale problems, especially in the context of iterative diagonalization methods.
    
    \item Robustness and good maintainability achieved via comprehensive unit testing (built on top of the Catch 2 framework \cite{Catch2}).
    
    \item Documentation covering all basic concepts of the library with detailed class and function descriptions and a set of usage examples.
\end{itemize}
It is worth noting that \libcommute\ does not provide any container types to store matrices, or algorithms to find their eigensystems. These tasks are left to the plentiful and well optimized C++ linear algebra libraries, such as Eigen \cite{Eigen} and Blaze \cite{Blaze}. Instead, \libcommute\ is focused on providing a convenient way to specify the relevant objects in the language of quantum operators.

\libcommute\ is accompanied by a set of Python 3 bindings called \pycommute. These are meant to further simplify development of ED codes in one the following ways.
\begin{itemize}
    \item \pycommute\ can be directly used to solve moderately sized problems when performance is not a concern.
    \item \pycommute\ is useful for rapid prototyping of \libcommute-based codes.
    \item \pycommute\ code may serve as a Python input layer for a high performance computational core written in C++ and wrapped in Python by means of pybind11 \cite{pybind11}.
\end{itemize}

\libcommute\ borrows and greatly extends upon some ideas from portions of the TRIQS library \cite{TRIQS}, which was co-written by the author of the present paper.
The $\pm h.c.$ syntax feature was borrowed from the excellent TBTK library \cite{TBTK} written and maintained by Kristofer Bj\"ornson. Another related work is the C++ port of the SeQuant library \cite{SeQuant2}, which features some facilities similar to those of \libcommute's Domain-Specific Language (see Sec.\ref{sec:expression} for details). Other open-source ED construction toolkits for C++ include EDLib \cite{EDLib}, pomerol \cite{pomerol}, parts of the ALPS library \cite{ALPS}, Qbasis \cite{Qbasis} and the 'lattice-symmetries' package \cite{lattice-symmetries}. The most notable Python projects in this category are QuSpin \cite{QuSpin1,QuSpin2}, PYED \cite{pyed}, pybinding \cite{pybinding} and KWANT \cite{KWANT} (the latter two are specialized for diagonalization of non-interacting tight-binding models).

\section{Software description}
\label{sec:description}

\subsection{Software Architecture}
\label{sec:architecture}

\libcommute\ is a C++11/14/17 template library that includes two major parts.

\begin{itemize}
    \item A Domain-Specific Language (DSL) designed to easily construct and manipulate polynomial expressions built out of quantum-mechanical operators used in the quantum many-body theory (fermion and boson creation and annihilation operators, spin ladder operators). The most commonly used instances of such expressions are many-body Hamiltonians and operators representing physical observables.

    \item A representation of the polynomial expressions in a form of linear operators, which enables their action on state vectors in finite-dimensional Hilbert spaces. This feature, together with a set of supplementary tools, provides a basis for writing highly performant ED codes without loss of flexibility.
\end{itemize}

\pycommute\ \cite{pycommute} is a Python package that uses pybind11 \cite{pybind11} to wrap a subset of \libcommute's features. It also provides an extra module for easy construction of some model Hamiltonians widely used in the theory of quantum many-body systems, quantum optics and the theory of spin lattices. \pycommute\ is available from the Python Package Index (PyPI) \cite{pycommute_pypi} and as part of a public Docker image \cite{pycommute_dockerhub}.

\subsection{Software Functionalities}
\label{sec:functionalities}

\subsubsection{The Domain-Specific Language}
\label{sec:expression}

\libcommute's DSL revolves around the notion of a polynomial expression
(the \verb|expression| class template). 
\begin{equation}\label{expression}
    E = C + \sum_\alpha C_\alpha g_\alpha
          + \sum_{\alpha\beta} C_{\alpha\beta} g_\alpha g_\beta
          + \sum_{\alpha\beta\gamma} C_{\alpha\beta\gamma}
                g_\alpha g_\beta g_\gamma + \ldots
\end{equation}

Expressions are assembled from non-commuting algebra generators $g_\alpha$ (such as fermion creation/annihilation operators $c^\dagger_{i,j,k,\ldots}, c_{i,j,k,\ldots}$) and numerical coefficients $C_{\alpha\beta\ldots}$ using a syntax that closely resembles mathematical notation. They support basic arithmetic operations $+$, $-$, $\cdot$ with simplification of their results, Hermitian conjugation, and the $\pm h.c.$ shorthand (plus/minus Hermitian conjugate). The operator-valued expressions are internally stored in the explicit form (\ref{expression}) rather than in a matrix representation, which means they do not suffer from the exponential dimensionality explosion, and their allowed degree is practically unlimited. Individual terms and generators in (\ref{expression}) are accessible via constant STL-like iterators.

Algebra generators carry statically typed lists of indices $i,j,k,\ldots$ which can represent integer lattice site coordinates, quasi-momentum components, spin or orbital string labels, or indices of a user-defined type. With C++17 available, it is also possible to construct expressions from generators with types of indices defined at runtime via the specially provided index type \verb|dynamic_indices|.

\libcommute\ supports building expressions out of fermionic ladder operators $c, c^\dagger$, bosonic operators $a, a^\dagger$ (generators of a canonical (anti)commutation relation algebra), and arbitrary (half-)integer spin operators $S_z, S_\pm$. Furthermore, generators of new algebras can be defined by deriving from the abstract base \verb|generator| and implementing a few methods that describe (anti)commutation relations and simplification rules for the new algebra\footnote{The online documentation contains examples on how to implement the Virasoro algebra and the algebra of Dirac $\gamma$-matrices.}.
In general, \libcommute's DSL can work with algebras, whose generators $g_\alpha$ obey
the commutation relations
\begin{equation}
    g_\alpha g_\beta - c g_\beta g_\alpha = F_{\alpha\beta} +
        \sum_\gamma f_{\alpha\beta}^\gamma g_\gamma
\end{equation}
with real constants $c$, $F_{\alpha\beta}$ and $f_{\alpha\beta}^\gamma$. Such algebraic structures include the Lie and Clifford algebras.

Another customization point for the DSL is the type of the numerical coefficients $C_{\alpha\beta\ldots}$. Beside the common double precision real and complex numbers, one can use a custom numeric-like type. This feature can prove useful when working with time-dependent Hamiltonians. Coefficients of the corresponding expressions can be number-like objects storing interpolators or power series w.r.t. the time variable. Specializing the \verb|scalar_traits| structure for the new coefficient type will teach \libcommute\ how to use it in an expression.
Mathematically speaking, any type whose values form a ring with operations $+$ and $\cdot$ can serve as a coefficient type (in fact, a ring without a multiplicative identity would suffice).

\subsubsection{Linear operators and Exact Diagonalization tools}
\label{sec:loperator}

The second major part of \libcommute, which makes it practically useful for writing exact diagonalization solvers, is the linear operator framework. Instances of the \verb|loperator| class represent action of polynomial expressions (\ref{expression}) on state vectors in finite dimensional Hilbert spaces. Such spaces are described by \verb|hilbert_space| objects and constructed as ordered direct products of elementary spaces $\mathcal{H}_i$,
\begin{equation}\label{hilbert_space}
    \mathcal{H} = \mathcal{H}_1 \otimes \mathcal{H}_2 \otimes \ldots \otimes
        \mathcal{H}_N.
\end{equation}
Each elementary space $\mathcal{H}_i$ corresponds to a single quantum degree of freedom (DOF) and can be one of the following.
\begin{itemize}
    \item A 2-dimensional space spanned by fermion occupation number states $|0\rangle$ and $|1\rangle$ (class \verb|elementary_space_fermion|).
    \item A truncated $2^b$-dimensional space spanned by boson occupation number states $|0\rangle, |1\rangle, \ldots, |2^b-1\rangle$. In particular, a $b = 1$ space corresponds to a hardcore boson DOF (class \verb|elementary_space_boson|).
    \item A $(2S+1)$-dimensional space corresponding to a spin-$S$ DOF (class \verb|elementary_space_spin|).
    \item An elementary space corresponding to a DOF from a user-defined algebra. The respective C++ types are classes implementing the \linebreak\verb|elementary_space| interface.
\end{itemize}
\verb|hilbert_space|'s API allows to explicitly build $\mathcal{H}$ out of \verb|elementary_space_*| objects corresponding to the factors $\mathcal{H}_i$. Another option is to delegate this task to the function \verb|make_hilbert_space()|, which automatically analyzes a polynomial expression and builds a minimal space required to accommodate the corresponding operator.

Instances of \verb|loperator| are callable C++ objects that can be applied to a state vector object -- a container or view type modeling a special \verb|StateVector| concept. \libcommute\ directly supports \verb|std::vector| and some vector-like types of the Eigen 3 library \cite{Eigen} as state vector objects, including the \linebreak\verb|Eigen::Map| views. Types provided by other matrix-vector algebra libraries can readily be made \verb|loperator|-compatible by overloading a few free functions for them. Requirements of the \verb|StateVector| concept are rather lax and can be met even by such intricate data types as GPU memory-based \cite{Thrust,BoostCompute,ArrayFire} and distributed arrays \cite{TrilinosPetra,DASH,GlobalArrays,upcxx}.

Following the same design principle as \verb|expression|, \verb|loperator| does not internally store any matrices (neither dense nor sparse) and describes action on the state vectors only as a set of element selection and transformation rules. This solution is memory efficient and fits particularly well with widely used iterative methods of partial diagonalization, such as power iteration \cite{PowerIteration} and the Lanczos algorithm \cite{ARPACK,ezARPACK,PRIMME}.

In addition to the basic building blocks of ED solvers -- Hilbert spaces, state vectors and linear operators -- \libcommute\ ships a handful of supplemental tools. \verb|sparse_state_vector| is a lightweight sparse array type that can be used to implement iterative ED solvers with repeated elimination of negligible quantum amplitudes \cite{HaverkortSolver}. \verb|mapped_basis_view| is a view of an underlying state vector container. It translates basis states of a Hilbert space according to a predefined map and is useful when only a specific sector of a model (block of its Hamiltonian) is to be diagonalized. A more specialized and less memory consuming version of such views is called \verb|n_fermion_(multi)sector_view| and works for the $N$-particle sectors of models with multiple fermionic DOF.
Finally, the \verb|space_partition| utility class reveals invariant subspaces of Hamiltonians using the automatic space partition algorithm described in Section 4 of \cite{CTHYB}. The algorithm reorders basis states in such a way that the Hamiltonian is brought to a block-diagonal form with the smallest possible block sizes.
It generates input basis state maps for \verb|mapped_basis_view| and, therefore, makes it possible to diagonalize individual sectors without knowing integrals of motion of the studied model.

\subsubsection{\pycommute}
\label{sec:pycommute}

The Python package \verb|pycommute| features three modules.
\begin{enumerate}
    \item \verb|expression| wraps the polynomial expression C++ types with real (class \verb|ExpressionR|) and complex coefficients (\verb|ExpressionC|),
    their arithmetics and functions to create generators of fermionic, bosonic and spin algebras. Support for user-defined algebras is planned for a future release.
    \item \verb|loperator| contains wrappers of Hilbert space types and linear operators acting on one-dimensional NumPy \cite{NumPy} arrays. It also exposes some of the state vector view types, the automatic space partitioning utilities, and a few overloads of the function \verb|make_matrix()|, which returns a matrix representation of a linear operator in a form of a two-dimensional NumPy array.
    \item \verb|models| is a set of factory functions for expressions of some widely used model Hamiltonians. These include tight-binding models, fermion and boson interaction terms, Ising, Heisenberg and Dzyaloshinskii–Moriya spin coupling terms, and a few spin-boson type models.
\end{enumerate}

It is possible to pass expressions, linear operators and other \pycommute\ objects between Python scripts and C++ functions defined in a pybind11-module.
This makes coding up interface layers of hybrid C++/Python ED codes a hassle-free task.

\section{Illustrative Examples}
\label{sec:examples}

This section briefly highlights a few \libcommute/\pycommute\ usage examples. Complete code listings of these and further examples with in-depth explanations are to be found on the respective documentation web-sites.

\subsection{\libcommute\ examples}
\label{sec:libcommute_examples}

The first example \cite{example_heisenberg_chain} concerns an integrable quantum system, a spin-$1/2$ Heisenberg chain. \libcommute's DSL is used to verify some explicit expressions for higher charges of the model derived in \cite{heisenberg_chain}.

The second example \cite{example_hubbard_holstein_1d} demonstrates how to partially diagonalize a one-dimensional Hubbard-Holstein model \cite{HubbardHolstein1,HubbardHolstein2} using \libcommute\ and Eigen 3 \cite{Eigen}. The model describes behavior of strongly correlated electrons on a lattice coupled to localized phonons. Diagonalization is performed within a sector with a fixed number of electrons.

\subsection{\pycommute\ examples}
\label{sec:pycommute_examples}

Both \pycommute\ examples mentioned below are simple Python scripts that make use of functions from the \verb|models| module to readily construct Hamiltonians, and of NumPy \cite{NumPy} to diagonalize them.
\begin{itemize}
    \item \cite{example_space_partition}: Sector-wise diagonalization of an interaction term of electrons in an atomic $d$-shell parametrized by Slater integrals $F^k$\cite{Slater,Judd}.
    \item \cite{example_tavis_cummings}: Spectrum of a generalized Jaynes-Cummings (Tavis-Cummings) model \cite{tavis_cummings} describing dynamics of two qubits coupled via an electromagnetic mode in a cavity.
\end{itemize}

\section{Impact}
\label{sec:impact}

The main anticipated impact of the presented software is a drastic reduction of development time for Exact Diagonalization codes written in C++ and (optionally) using Python scripting as their input/output layer. \libcommute's domain specific language provides a very convenient and flexible way to deal with quantum-mechanical operators, while the \verb|loperator| module abstracts out a lot of small and hard-to-keep-track-of details. Availability of an open source, well tested, well documented and highly reusable framework -- such as \libcommute\ -- can make building ED codes a much less tedious and error-prone task. Since \libcommute\ has been and will continue to be developed with speed and memory consumption in mind, its users can enjoy a decent performance level without investing much time into profiling and fine tuning.

The development time can be further reduced by first trying out new ideas on the level of \pycommute-based Python scripts. As \libcommute\ and \pycommute\ share the same set of basic concepts, translating the scripts into HPC-ready C++ code is fairly easy.

Last but not least, \pycommute\ allows one to define Hamiltonians and operators of physical observables and pass them to various \libcommute/\pycommute-based solvers in a uniform manner. It can, therefore, establish a universal input data format for such computational programs.

\libcommute/\pycommute\ is still a very young project. At the moment, it provides a foundation for the recent 2.0 release of the Pomerol ED library \cite{pomerol}, as well as for a few privately developed solvers. Nonetheless, wider adoption of the framework by the community would certainly simplify development of modern and complex ED solvers, while improving reliability and reproducibility of their results.

\section{Conclusions}
\label{sec:conclusions}

To conclude, I have presented \libcommute/\pycommute, a combination of a C++11 template library and its Python bindings, whose primary goal is rapid development of complex Exact Diagonalization solvers for models of quantum many-body theory. Despite being a young project, it already has the potential to provide a solid foundation for future high performance scientific software, which is both easy to write and to use. Future development of \libcommute\ will focus on expanding the arsenal of the ED construction tools. Newer versions of \pycommute\ are going to expose more of \libcommute's functionality, in particular, make it possible to define new operator algebras in the Python code.

\section{Conflict of Interest}
\label{sec:conflict}
No conflict of interest exists:
I wish to confirm that there are no known conflicts of interest associated with this publication and there has been no significant financial support for this work that could have influenced its outcome.

\section*{Acknowledgements}
\label{sec:acknowledgements}

I would like to acknowledge contributions from Xinyang Dong, who provided useful feedback at the early stages of project's development, and from Hugo U. R. Strand for his feature suggestions. I am also grateful to Joseph Kleinhenz for proofreading this manuscript.
This research did not receive any specific grant from funding agencies in the public, commercial, or not-for-profit sectors.

\bibliographystyle{elsarticle-num}
\bibliography{main}

\end{document}